# Kerr-microresonator solitons for accurate carrier-envelope-frequency stabilization

T. C. Briles[1,2,*], J. R. Stone[1,2], T. E. Drake[1], D. T. Spencer[1], C. Frederick[1,2], Q. Li[3], D. A. Westly[3], B. R. Illic[3], K. Srinivasan[3], S. A. Diddams[1,2], and S. B. Papp[1,2]

**Affiliations:**

[1]Time and Frequency Division, National Institute of Standards and Technology, Boulder, Colorado 80305, USA.

[2]Department of Physics, University of Colorado, Boulder, Colorado, 80309, USA.

[3]Center for Nanoscale Science and Technology, National Institute of Standards and Technology, Gaithersburg, Maryland 20899, USA.

*travis.briles@nist.gov

**Carrier-envelope phase stabilization of optical pulses enables exquisitely precise measurements by way of direct optical-frequency synthesis, absolute optical-to-microwave phase conversion, and control of ultrafast waveforms. We report such phase stabilization for Kerr-microresonator frequency combs integrated on silicon chips, and verify their fractional-frequency inaccuracy at <3x10$^{-16}$. Our work introduces an interlocked Kerr-comb configuration comprised of one silicon-nitride and one silica microresonator, which feature nearly harmonic repetition frequencies and can be generated with one laser. These frequency combs support an ultrafast-laser regime with few-optical-cycle, 1-picosecond-period soliton pulses and a total dispersive-wave-enhanced bandwidth of 170 THz, while providing a stable phase-link between the optical and microwave domains. To accommodate low-power and mobile application platforms, our phase-locked frequency-comb system operates with <250 mW of chip-coupled power. Our work establishes Kerr-microresonator combs as a viable technology for applications like optical-atomic timekeeping, optical synthesis, and related directions.**

The invention of optical frequency combs (*1*, *2*) opened many new applications (*3*) in photonics from precision timing and ranging to generation of entangled states (*4*, *5*). They are composed of hundreds to millions of optical modes whose frequencies conform to a simple relationship, $v_n = n \times f_{\text{rep}} + f_0$, where $n$ is the mode number, $f_{\text{rep}}$ is the repetition frequency, and $f_0$ is the carrier-envelope-phase offset frequency. A consequence of this basic equation is that complete knowledge of the comb lines can be measured or electronically synthesized (*6*) with practically arbitrary precision. Phase locking $f_{\text{rep}}$ and $f_0$ delivers to users a fully stabilized frequency comb, which can be derived from the SI second or another reference. Such phase stabilization is what sets optical-frequency combs apart from other laser frequency measurement tools that require post-processing or measurement calibration.

Recently, investigations of Kerr microresonators have led to advances in parametric-optics comb generation (*7*), particularly in ultrahigh-speed optical waveforms (*8*) (*9*) and photonic-chip integration (*10*). Dissipative-Kerr soliton (DKS) pulses that stably propagate in Kerr microresonators (*11*, *12*) are thus far the most useful localized nonlinear pattern of the electromagnetic field. Owing to the unique role of the input CW laser at frequency $v_p$, the mode frequencies of DKS combs are conveniently expressed in terms of the mode number, µ, relative to the pump laser, according to $v_\mu = v_p + \mu \times f_{\text{rep}}$. There have been several advances in the field, including development of various resonator materials (*10*, *13*, *14*); demonstrations of visible

(*15*, *16*), mid-infrared (*17*, *18*), and octave-spanning operation (*19*, *20*); chip-based optical clocks (*21*); integration into complex photonic circuits (*22*); and studies on DKS breathing (*23*, *24*) and crystallization (*25*). DKS combs have also been developed for the goal of 2f-3f and f-2f self-referencing with external spectral broadening for counting optical cycles (*26*), carrier-offset frequency control (*27*), and optical-clock comparisons (*28*); and for 2f-3f measurements with a photonic-chip-integrated soliton comb (*29*). Still, Kerr combs have yet to see the same kind of $f_0$ stabilization as is routine for modelocked-laser combs.

In this work, we introduce a Kerr-frequency-comb system built around $f_0$ phase stabilization, realizing this technique entirely with microresonators for the first time. We draw on notable strengths of Kerr-soliton combs – low noise, a wide design space, and chip integration. Our interlocked configuration of one silica and one silicon-nitride (SiN) resonator leverages both nanofabrication and photonic-chip integration. Operating the silica Kerr comb at a microwave rate provides a dense but narrow-bandwidth reference grid, while the terahertz-rate SiN comb provides more than an octave of bandwidth for f-2f self-referencing. The key to $f_0$ stabilization is design and control of the ultrafast pulse circulating in the SiN resonator, including its multiple quasi-CW dispersive waves (DWs) and the magnitude of $f_0$ that can vary practically to 0.5 THz. Our work explores interesting regimes in ultrafast photonics, highlighted by our observation of self-interaction effects in which the full envelope of a three-optical-cycle soliton, including the radiation tails from the DWs, extends entirely around the microresonator. Still, we show that such a complex soliton Kerr comb permits low-noise phase stabilization of its repetition and offset frequencies with respect to the SI second, and we verify its accuracy and precision to better than 3x10$^{-16}$ through out-of-loop measurements with an auxiliary self-referenced comb.

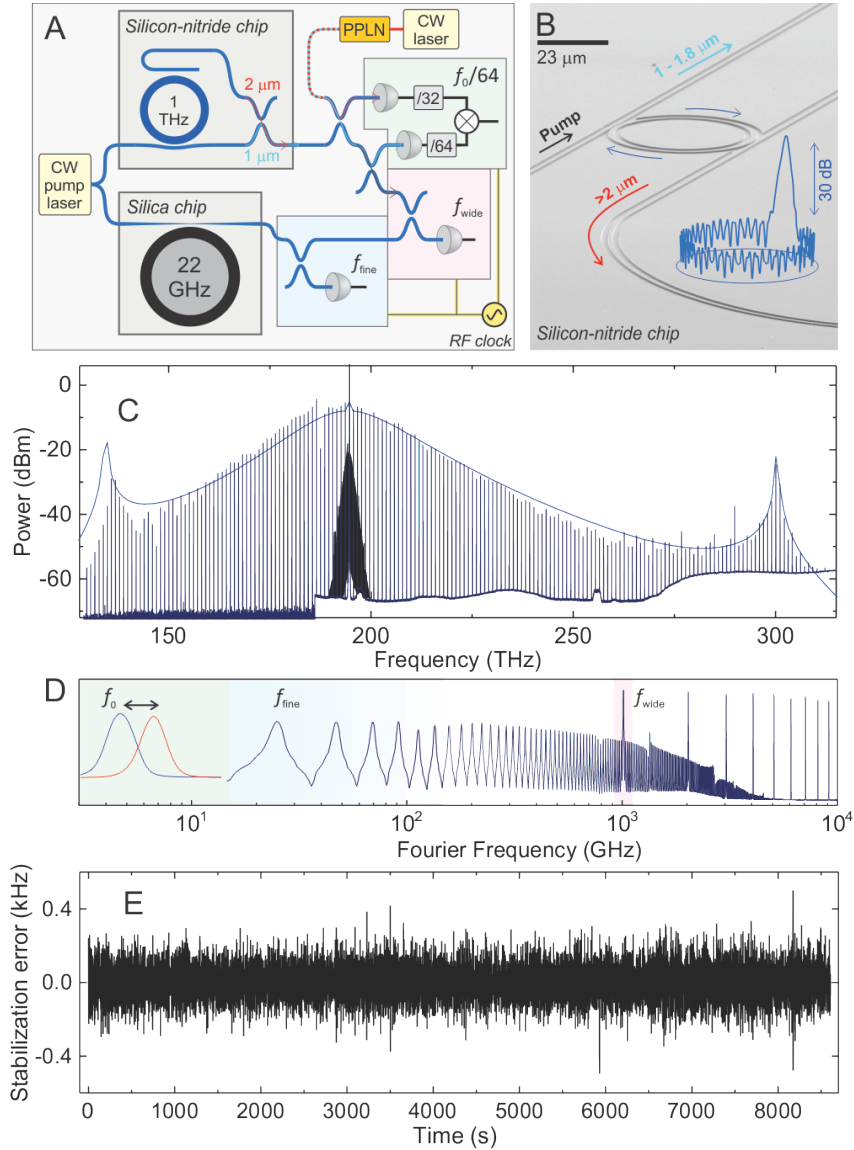

Figure 1: Interlocked Kerr comb for absolute $f_0$ stabilization. (A) System diagram indicating the microresonators, photonics, the detection electronics, and RF clock for stabilization. Frequency mixing as indicated yields $f_0$ in our experiments. PPLN: Periodically poled LiNbO$_3$. (B) Scanning-electron microscope (SEM) image of the SiN chip with add/drop coupler configuration. (C) Optical spectrum of the two-soliton comb. The 1-THz-spaced modes are easily resolved, and the silica comb modes could be seen by expanding the plot between 190 THz and 200 THz. (D) Conceptual hierarchy of the optical-frequency-synthesis scheme exploiting stepwise up-conversion of an input RF clock. The highlighted frequency ranges indicate electronically detectable $f_0$, and fine and wide repetition frequencies, $f_{\text{fine}}$ and $f_{\text{wide}}$, respectively. (E) Absolute optical stabilization error of the interlocked Kerr combs, evaluated by comparison to an auxiliary self-referenced erbium-fiber frequency comb. The frequency counter gate time is 1 second.

Figure 1 presents an overview of our experiment and key results, and the subsequent figures present analysis of the system. We implement an interlocked, dual-microresonator Kerr comb (Fig. 1A), which is designed principally for photonic-chip integration and low-power consumption. When interlocked with the silica comb, the SiN comb mode frequencies are $v_n = n\,f_{\text{wide}} + f_0 = n\,(H\,f_{\text{fine}} + f_{\text{rem}}) + f_0$, where $f_{\text{wide}}$ ($f_{\text{fine}}$) is the repetition frequency of the SiN (silica) comb, and H and $f_{\text{rem}}$ are the integer part and remainder of the ratio $f_{\text{wide}}/f_{\text{fine}}$. A similar algebraic relationship exists for the full interlocked DKS. The near-harmonic relationship between $f_{\text{wide}}$ and $f_{\text{fine}}$ optimizes the threshold powers of our combs, which vary inversely with microresonator free-spectral ranges (FSR), and can be equivalent due to the silica microresonator having a factor of 100 greater quality factor ($Q$). The silica chip contains a wedge resonator with 21.97 GHz FSR that has been described elsewhere (*31*, *32*); we use a tapered fiber to couple up to 50 mW to the device. The SiN chip is fabricated on a thermally-oxidized silicon wafer (oxide thickness is 3 µm); a ~615 nm +/- 2 nm thick low-pressure, chemical-vapor deposition (LPCVD) stoichiometric silicon-nitride ($Si_3N_4$) film is patterned by electron-beam lithography and dry etching to create a 23 µm radius ring resonator and two integrated waveguide couplers. The top cladding is air for broadband resonator dispersion optimization (*19*). Output coupling the >180 THz comb bandwidth supported by the ring resonator in this geometry requires a careful design: A pulley coupler (*33*) provides $Q_{\text{ext}} < 10^6$ for most of the 900-1900 nm wavelength range and supports pumping in the 1550 nm band with 200 mW of on-chip power; a straight waveguide provides efficient output coupling as a drop port beyond 1900 nm; and an on-chip dichroic coupler efficiently combines longwave light from the drop port into the access waveguide for extraction of the entire comb spectrum from a single waveguide. Figure 1B presents an SEM image of the SiN chip with annotations indicating the role of the two couplers. Further coupling details are in the supporting material (*34*).

Operationally, we generate DKS pulses in both microresonators by suitable frequency control of the pump laser; an optical spectrum of the interlocked soliton comb is shown in Fig. 1C. The finely spaced silica comb covers the telecom C-band, and the SiN comb with corresponding sech$^2$ DKS duration of 3 optical cycles extends over more than 170 THz across the near IR, including its two DW components at 300 THz (999 nm) and 137 THz (2191 nm) that arise from higher-order corrections to the group-velocity dispersion (*19*, *35*, *36*). The blue trace results from a simulation of the Lugiatio-Lefever equation (LLE) (*12*) that mostly captures the spectral envelope of the SiN soliton comb through our measurements of $Q$ and pump power, and finite-element modeling (FEM) of the device mode frequencies. The small discrepancies between the observed and simulated spectra are attributed to a ~5 nm precision in dimensions of the fabricated resonator, the dimensions used in the FEM simulation, and residual wavelength dependent absorption in the SiN material. The corresponding LLE intracavity intensity is inset to Fig. 1B and shows the few-optical-cycle DKS propagating atop the CW pump background and the dispersive waves.

The goal of our work is to photodetect and electronically phase lock the three degrees of freedom ($f_0, f_{\text{wide}}, f_{\text{fine}}$) that define the $v_n$ mode spectrum of the combs. To understand their relationships, Fig. 1D presents a Fourier-domain view of the frequency combs that would notionally be created by photodetection of the comb after second-harmonic generation for f-2f self-referencing. This plot highlights how all the optical comb modes – and the 12 GHz $f_0$ frequency of the wide comb – are resolved by a moderate resolution spectrometer. Activating phase-locked loops for ($f_0$, $f_{\text{wide}}$, $f_{\text{fine}}$) with respect to a hydrogen-maser RF clock (Fig. 1A) fully stabilizes the combs. We perform an out-of-loop verification of the interlocked Kerr comb using an auxiliary reference comb, which is also stabilized to the RF clock. Figure 1E shows a nearly 9000 s frequency-counter record of the optical heterodyne beat between the two combs, indicating their relative statistical fluctuations of 117 Hz in a 1 s observation window and mean difference of $-0.052 \pm 0.057$ Hz on the 194.635,926,272,000 THz carrier frequency.

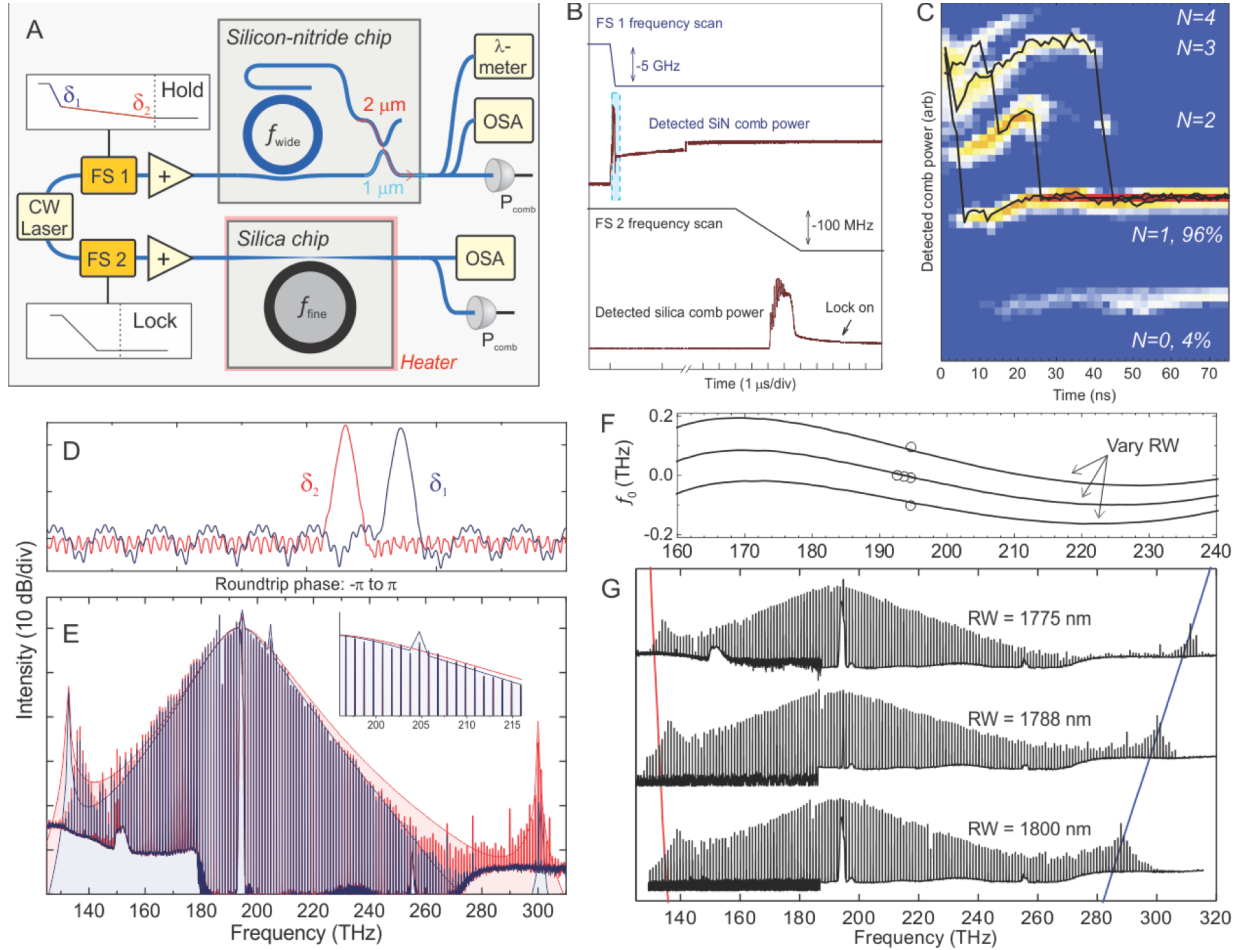

Figure 2: Interlocked Kerr comb formation. (A) Apparatus and diagnostics for comb formation. (B) Pump-laser-frequency sweeps and observed comb power as solitons emerge from the chaotic MI regime. (C) Histogram plot of SiN comb formation involving multiple soliton number labeled N, our desired N=1 state, and an empty cavity. Three representative traces (black lines) of single experiments. Here the photon lifetime is 0.3 ns. (D, E) Observations of destructive self-interaction between mode-interaction-induced excess comb light and DW components of the soliton waveform at resonator detuning $\delta_1$ (blue) and $\delta_1$ (red). (D) The simulated pulse envelope and modulation of the background intracavity power from the mode interaction at 205 THz. (E) The measured optical spectra and corresponding LLE simulation results for the spectral envelope. (F, G) Design and control of $f_0$ and the SiN soliton bandwidth through the device RW. Curves in (F) are ordered as in (G). The solid red and blue lines track the expected locations of the dispersive waves from the LLE simulation.

Figure 2 shows how we realize the interlocked Kerr combs. We initiate DKS in both devices by a programmed sequence of pump-laser frequency sweeps (*34*). An external cavity diode laser (Fig. 2A) provides slow frequency tuning, and dual-Mach-Zehnder-intensity-modulator optical-frequency shifters (FS 1 and FS 2) enable individualized rapid sweeps for the microresonators. A heater adjusts the silica chip to align the resonant frequencies of the SiN TE1 and silica TM1 modes near 1540 nm. We diagnose soliton formation by photodetection of the comb intensity; Fig. 2B shows sweep timings and comb-power traces as the combs discontinuously evolve from the modulation-instability (MI) regime to Kerr solitons. To initiate the SiN soliton, we find it adequate to program an open-loop (*37*), two-segment sweep first at 5 GHz per 100 ns to a particular resonator detuning $\delta_1$ (*34*), and then at 1 GHz/s to a larger effectively red detuning $\delta_2$. For the silica comb, we implement an offset Pound-Drever-Hall lock via FS 2 to maintain the soliton as described in Ref. (*38*). In essence, this procedure reduces thermo-optic resonance shifts of the devices (*11*), and enables us to stabilize single solitons with one (or two for convenience) tunable pump laser(s). We study the ns-scale dynamics of SiN soliton formation by performing 500 trials in 5 ms; Fig. 2C presents a logarithmic density histogram of comb power in the soliton regime, which correlates to soliton number. The plot begins within ±3 ns of reaching the soliton regime and ends at the observed steady-state condition. Here the sweep rate to $\delta_1$ largely controls the soliton number distribution and we optimize it for generation of single soliton pulses, reaching a fidelity of 96%. In the tens of ns following soliton formation, we observe interesting soliton number transitions, mostly in units of one, as the intracavity waveform stabilizes after the MI transition. In the future, this may provide a unique capability for comparisons with LLE dynamical simulations.

Our experiments require not only DKS generation in the SiN device, but formation of intense dispersive-wave enhancements to aid f-2f interferometry, and adjustment of $f_0$ to less than ~20 GHz for practical photodetection. Here we present our experimental characterization of this optimization procedure. Figures 2D and 2E explore the evolution of the soliton comb spectrum as we tune the pump laser between $\delta_1$ and $\delta_2$. Unexpectedly, we observe a sudden increase of 20 dB in the 300-THz DW intensity, as indicated by the difference in blue and red comb-spectra traces shown in Fig. 2E. Since this phenomenon represents a comparable 20 dB increase in the signal-noise ration (SNR) of $f_0$ and practically determines whether $f_0$ can be detected, we search in detail for its cause. By monitoring the soliton spectrum as we tune between $\delta_1$ and $\delta_2$, we first observe a strong mode interaction that aberrantly enhances the ~205 THz comb-mode intensity. As the detuning is increased further to $\delta_2$, the mode interaction diminishes towards the sech$^2$ profile followed by a sudden increase of the 300-THz DW (*34*). Hence, we observe that DW formation with single Kerr solitons is sensitive to microresonator mode structure perturbations. Within the framework of a mode-interaction-perturbed LLE (*25*, *39*, *40*), we simulate that DW formation is suppressed by the extended background wave from excess comb light at 205 THz. The LLE simulations in Fig. 2D capture our observations by inclusion of a mode interaction; see the inset of Fig. 2E. Since both the DW and the extended background wave wrap around the SiN microresonator in both space and time there is the opportunity for interference. The strong reduction of the DW intensity implies that the interference is destructive, providing evidence for a self-interaction regime in a resonator with one Kerr soliton in which DW generation competes with other quasi-CW components of the intracavity field. Interestingly, our experiments and simulations agree that the DW intensity at 137 THz is largely unaffected by the 205 THz mode interaction.

We adjust $f_0$, the soliton bandwidth, and the dispersive-wave spectral peaks by optimizing the geometry of the SiN ring. The ring radius (RR) and the ring-waveguide width (RW) are varied in the nanofabrication process through different device instances on the same chip. Figure 2F and 2G present an investigation into how three discrete 12.5 nm adjustments to RW affect the Kerr-soliton comb $f_0$ frequency and DW wavelengths, respectively. In particular, the carrier-envelope-offset frequency is $f_0 = \nu_p - M f_{\text{wide}}$, where $M$ is the mode number of the pump mode in our experiments. We characterize $f_0$ at the MHz level with a wavemeter measurement of $\nu_p$ and knowledge of $f_{\text{wide}}$, which depends sensitively on RW and RR, and on the mode number of the pumped resonance, since GVD is wavelength dependent. These behaviors are captured in Fig. 2F in which the RW is varied, and we measure $f_0$ for three adjacent modes at 1540 nm, 1548 nm, and 1556 nm for RW=1787.5 nm. Agreement with FEM simulations of the SiN TE1 mode frequencies (black lines) is critical in closing the loop between design, nanofabrication, and operation of comb systems. Our present analysis requires a correction of only ~10 nm to the SiN dimensions used in the FEM analysis. The soliton bandwidth and DW wavelengths also vary significantly with RW (Fig. 2G), and this is critical for matching the Kerr-comb spectrum to applications. The FEM results used in the LLE dispersive-wave predictions (Fig. 2G) are accurate enough that a sweep of a few SiN rings on a single chip can yield the desired spectral coverage.

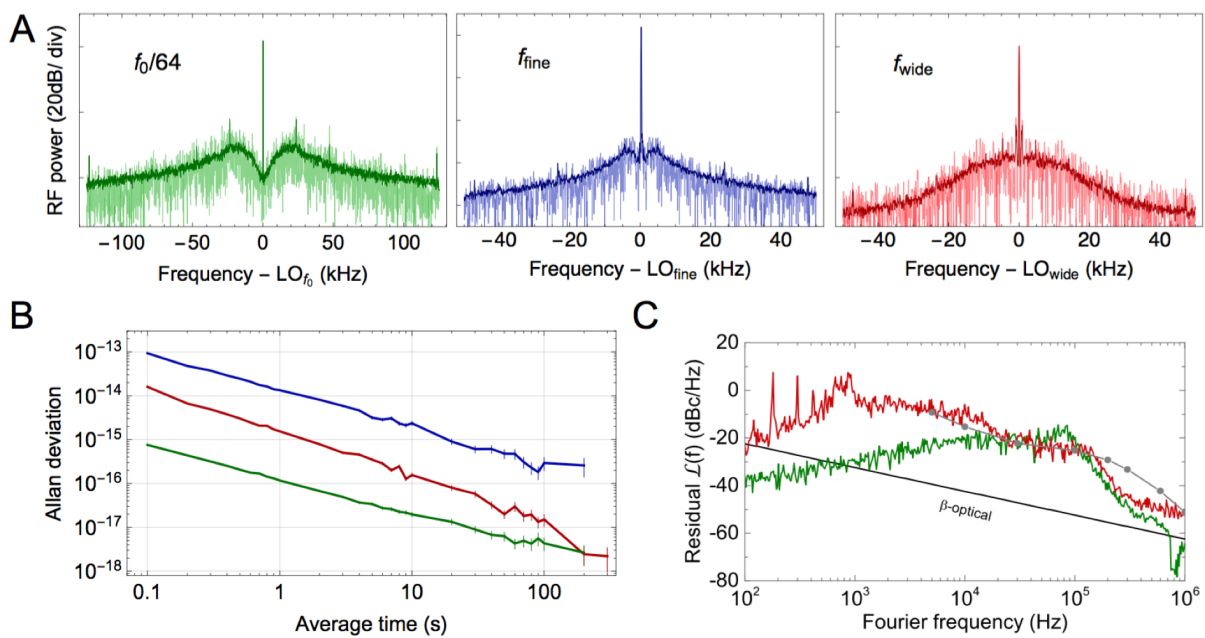

Figure 3: Phase stabilization of the interlocked Kerr combs. (A) Power-spectral density of the carrier-envelope-offset frequency (green), the silica comb repetition frequency (blue), and the SiN comb repetition frequency (red). Each panel displays single acquisition (lighter shade) as well as a trace with 25 avgs (darker shade) (B) Residual-frequency-noise contribution of the phase locks. Frequency counter gate time is 0.1 s and normalization is to the carrier frequency of the signals. (C) Phase noise of $f_{wide}$ (red) and $f_0$ (green), and the expected contribution from a microwave synthesizer (gray). Noise components above the $\beta$-line contribute to frequency comb linewidth.

With the interlocked Kerr combs of Fig. 1C, we focus on their phase stabilization (Fig. 3A) with respect to the RF clock. Due to off-chip coupling losses, somewhat insufficient SiN comb line powers, and the second-harmonic generation (SHG) efficiency of available PPLN devices at 150 THz (1998 nm), we utilize an auxiliary CW laser operating near the THz comb tooth at 150 THz. The relative detuning between the SiN comb and SHG of CW laser at 150 THz (300 THz) is monitored with optical-heterodyne signals, which are electronically processed to generate a single RF tone that is exactly $f_0/64$ and subsequently stabilized in a PLL (green box of Fig. 1A). The silica comb repetition frequency is photodetected; see blue box of Fig. 1A. The small remainder, $f_{\text{rem}}$, is measured by taking a heterodyne between the two combs, which is then electronically processed with the 46[th] harmonic of $f_{\text{fine}}$ to generate $f_{\text{wide}}$; see the red box of Fig. 1A. To stabilize these three signals, we search for the optimum electronic feedback pathways to the combs, which we currently understand to be pump laser power for $f_0$ and the frequency shifter settings of FS 1 and FS 2 for $f_{\text{wide}}$ and $f_{\text{fine}}$, respectively. Digital phase locks are applied simultaneously, and Fig. 3A shows the power-spectral density of these three stabilized signals. The wide and fine comb repetition frequencies feature prominent signals at the lock points, while $f_0$ following digital division by 64 is coherently locked to its RF reference.

We characterize the residual-noise contribution of the phase locks, which is additive with RF-clock noise, to understand the absolute frequency stability of the interlocked comb modes. We record the phase-locked signals with a continuous Π-type frequency counter, which we convert to Allan deviation (Fig. 3B), and with a phase-noise analyzer (Fig. 3C). No more than 2 Hz of noise is contributed to the optical modes in a 1-sec acquisition time, and the contribution is less with longer acquisitions due to the relative coherence to the RF clock that also gates the frequency counter. Phase-noise spectra of the locked $f_{\text{wide}}$ and $f_0$ signals provide complementary information. Servo oscillations appear at the 1 kHz and 100 kHz loop bandwidths for $f_{\text{wide}}$ and $f_0$. Due to the near-harmonic relationship between them, the SiN comb is very sensitive to 22 GHz phase noise of the silica comb's microwave reference oscillator, which we indicate by the gray trace in Fig. 3C. In our current configuration using an RF clock, we would not increase the $f_{\text{wide}}$ servo bandwidth because the un-stabilized $f_{\text{wide}}$ noise is commensurate with high-performance microwave synthesizers. Further, we assess the degree of coherence between the comb and the reference oscillator that would be required to realize a narrow linewidth SiN Kerr comb that would enable ultraprecise measurements (*41*). In Fig. 3C we draw the $\beta$-separation line, which estimates laser linewidth contributions based on phase-noise measurements (*42*). The $\beta$ line is relatively simple to understand, and phase noise above it leads to increased linewidth. Given the microwave frequency reference used here, $f_{\text{wide}}$ is too noisy. Our current inability to passively, or through servo control, reduce the carrier-offset phase noise represents an outstanding challenge in Kerr combs compared to the most advanced tabletop combs. Still both signals permit highly coherent measurements with long acquisition times (*41*).

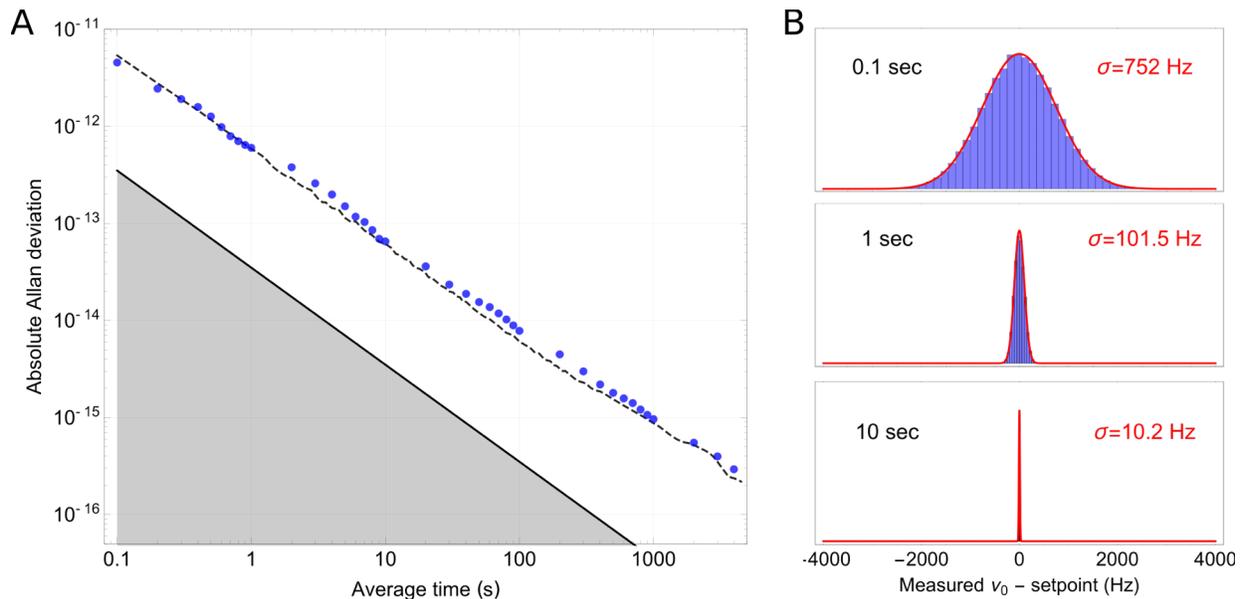

Figure 4: Stability and accuracy of the SiN $\mu = 0$ comb mode. (A) Allan deviation (blue points) measured with the auxiliary comb, and the total contribution from microwave synthesizers (dashed line). The gray shaded region indicates the residual noise of the Kerr-comb phase locks. (B) Histogram of frequency counter (0.1 s gate time) record used to obtain the overlapping Allan deviation for 0.1 sec, 1 sec, and 10 sec acquisitions.

Complete phase stabilization of our interlocked Kerr comb synthesizes an accurate and precise set of comb mode frequencies $v_n$ with respect to the RF setpoints of $f_{\text{fine}}$, $f_{\text{wide}}$, and $f_0$, which are all derived from the RF clock. For verification, we divert a fraction of the SiN comb output for frequency measurements via optical heterodyne beats with an auxiliary self-referenced erbium-fiber optical frequency comb (*43*) that is referenced to the RF clock. The result is a frequency-counter record of the pump laser SiN mode at 194 THz (Fig. 1E) that we analyze with the overlapping Allan deviation (Fig. 4A). Similar data and conclusions for the $\mu = (-1,1,2,3)$ SiN comb modes are presented in the supporting material (*34*), and we can infer the stability of all the comb modes, since f-2f stabilization depends on the entire comb span. The absolute fractional-frequency stability of the synthesized comb is $6 \times 10^{-13}/\tau$, where $\tau$ indicates the observed linear dependence with average time. This behavior confirms the stable phase lock and predominately white phase noise of the interlocked Kerr comb, the auxiliary comb, and the frequency counter with respect to the RF clock (*44*). Even though both combs share the same RF clock, they utilize separate electronic frequency synthesizers in their phase-locked loops. We measure the synthesizers' noise contribution (black dashed line in Fig. 4A) by frequency mixing them and counting their difference using the same acquisition mode of the frequency counter as for the optical-heterodyne beats. The synthesizers' noise contribution is currently the limiting factor in our optical synthesis chain. The residual noise added to the optical-comb frequencies by the feedback loops is estimated by adding the variances of the in-loop noise shown in Fig. 3B. An estimate of this noise is $4 \times 10^{-14}/\tau$ (solid black line in Fig. 4A) indicating the potential to utilize much more stable reference oscillators, such as all-optical locking.

A user can take advantage of phase-stabilized frequency combs for synchronization between remote locations, like time transfer and multi-static radar detection. We utilize a ~30-meter fiber path between laboratories without active noise cancellation to compare the interlocked Kerr comb and the auxiliary comb, and the combs' relative instability is determined by the microwave electronics involved. Still, the benefit of phase coherence between the combs can be understood through their heterodyne-beat frequency counter lineshapes; Fig. 4B presents a progression of histograms obtained by binning one record that indicates an optical-frequency measurement imprecision of only 10 Hz after 10 s of measurement time.

In conclusion, we have demonstrated carrier-envelope phase stabilization in Kerr microresonator combs, bringing ultraprecise optical-frequency synthesis to photonic-chip devices. Our work explores the generation of three-optical-cycle, ultrabroad bandwidth soliton pulses that repeat every 1 ps, and we leverage key operational advantages of Kerr microresonators such as small size, photonic integration, and low-power consumption. Stabilization of the combs synthesizes to the optical domain an RF clock that can be traced back to the SI second, enabling precision time and frequency metrology in a chip-scale package. With enhancements to the photonics packaging that we use, and targeted improvements to the SiN comb line power and SHG technology for f-2f detection, it appears possible to reduce our benchtop demonstration to a centimeter-scale stabilized frequency comb.

We thank Daniel Hickstein and David Carlson for helpful revisions on the manuscript, Kerry Vahala for the silica resonator, SriCo for the waveguide PPLN, and Tobias Kippenberg for loan of the auxiliary CW laser used for $f_0$ detection. We also acknowledge Robert Lutwak, the DODOS program management team, and our collaborators for helpful comments throughout the experiments. This work is funded by the DARPA DODOS program, NIST, and NRC. This work is not subject to copyright in the United States.

# Supplementary Materials: Kerr-microresonator solitons for accurate carrier-envelope-frequency stabilization

**Materials and Methods:** *Silicon nitride photonic circuit*

The photonic circuit surrounding the silicon-nitride resonator (Fig. 1B) is optimized for broad spectral bandwidth by addressing both the (1) wavelength dependent waveguide-resonator coupling rates and (2) the coupling of over 180 THz bandwidth into a single waveguide for convenient extraction off chip. A microscope image of the silicon-nitride chip highlighting three different devices is shown in Fig. S1A. The coupler design must achieve adequate coupling rates in the pump band for low power operation of the comb and also at edges of the spectrum for out-coupling of the dispersive waves. A conventional straight waveguide coupler optimized for a given spectral region, for instance near the 1.55um pump band, will typically be strongly undercoupled at short wavelengths and strongly overcoupled at longer wavelengths (*1*). With such a coupler, only a small fraction of the intracavity soliton spectrum would be efficiently outcoupled from the resonator to the waveguide. In our work, we use an add/drop configuration with a pulley coupler on the through port, which is optimized for the short wavelength dispersive wave and telecom C band and a straight waveguide coupler on the drop port which is optimized for extracting long wavelengths. Figure S1B shows a microscope image of one of the SiN resonators (middle panel) as well as enlargements of the coupling regions for the pulley coupler (top panel) and straight coupler (bottom panel.) The through port has a waveguide width of 720 nm, a pulley length of 4um, and a separation of 400 nm between the waveguide and resonator. The drop port has a waveguide width of 1200 nm and gap of 800 nm. The thickness of the SiN film is estimated to be $615 \pm 2$ nm. Figure S1C shows finite element simulations of the coupling Q of the through port (blue line) and drop port (red line) of the structure used in this work. As a result of its asymmetric cladding (top air cladding, bottom SiO2 cladding) and small cross-section, the through port waveguide has a cutoff wavelength near 1850 nm. The blue and purple shaded boxes indicate the wavelength of the 1 μm dispersive wave and pump laser. The red shaded box indicates the drop port for the >2 μm dispersive wave.

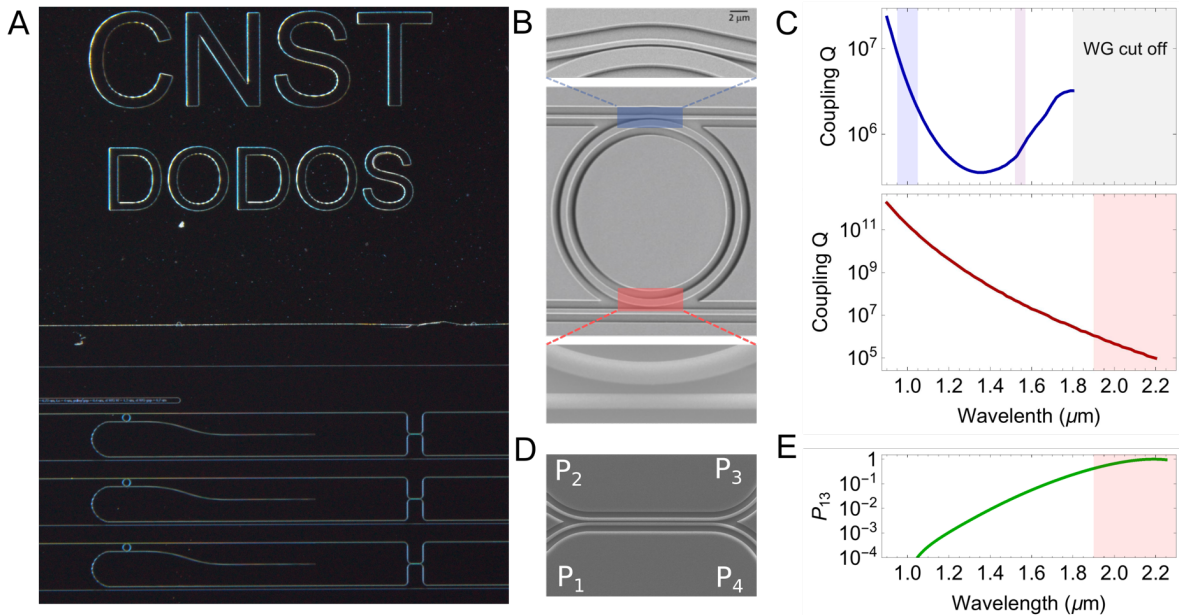

Figure S1: Design of the photonic circuit used to extract light from resonator and off of the chip. (A) Photograph of silicon nitride chip showing 3 different resonators with dichroic couplers. (B) SEM image of add/drop coupler configuration with pulley coupler on the through port and straight coupler on the drop port. (C) Simulated coupling Q of the pulley coupler (top) and straight coupler (bottom). For the pulley coupler, regions of efficient coupling of the short-wavelength dispersive wave and pump band are indicated with blue and purple boxes respectively. For the straight coupler, the region of efficient extraction of the long-wavelength dispersive wave is indicated with a pink box. (D,E) On chip dichroic coupler. (D) SEM image of the on chip dichroic coupler to transfer light from the drop port to the access waveguide. (E) Simulated power coupling ratio for light in the drop port to transfer to the through port in the dichroic coupler.

After extraction from the resonator, light in each of the two waveguides is combined in an on-chip dichroic coupler. A photograph of the coupler with annotations labeling each of the four ports is shown in Fig. S1D. The through port waveguide is adiabatically tapered from 720 nm to 1200 nm over a distance of 1 mm for phase matching with the drop port waveguide. The two waveguides are separated by a gap of 500 nm over a parallel interaction length of 45 μm. The simulated power coupling ratio, $P_{13}$, shown in Fig. S1E indicates that light at wavelengths >1900 nm is efficiently coupled from the drop port to the through port. Light is extracted off the chip via port $P_3$. By reciprocity, path $P_{24}$ has the same wavelength dependence as $P_{13}$ but does not constitute a loss channel due to the long wavelength cutoff in the pulley coupler.

**Supplemental Text**
*Soliton Generation in SiN resonators*

An essential ingredient of the demonstrated interlocked DKS system is the reliable generation of soliton combs in each microresonator. The key parameter governing the generation of solitons is the relative detuning between the pump laser and the thermally shifted cavity resonance (*2, 3*). Here, we define the pump-resonator detuning as $\delta = \nu_C(T) - \nu_p$, where $\nu_C(T)$ is the thermally shifted cavity resonance frequency at temperature T, and $\nu_p$ is the pump frequency. In this convention, $\delta > 0$ corresponds to the pump laser being effectively 'red' detuned relative to the thermally shifted cavity. In general, chaotic modulation instability (MI) exists at effective blue detuning, while mode-locked soliton states exist at effective red detuning. The transition from the chaotic region to the soliton region of phase space induces transient thermal shifts in the cavity which can lead to a loss of the soliton state. These thermal transients can be overcome with agile control of the pump-resonator detuning to ensure operation within the soliton existence range. Various approaches have been developed for dealing with the thermal transient based on control of the pump laser, including slow ramps of the pump laser frequency (*2, 4*), modulation of the pump power (*5, 6*), backwards tuning (*7*), direct modulation of the cavity resonance center frequency with integrated heaters (*8*), as well as indirect (*6*) and direct (*9, 10*) active stabilization of the detuning.

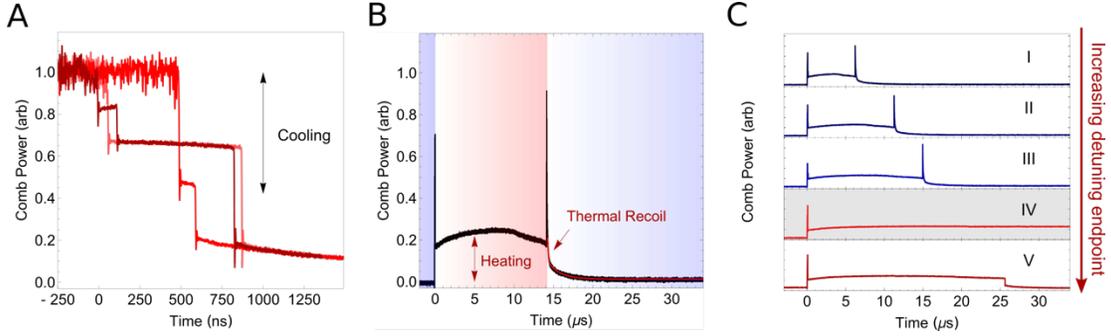

Figure S2: Soliton generation in SiN resonators with fast laser sweeps. (A) Thermal timescale limited soliton steps generated by slow sweeps across resonance. (B) Conceptual overview of soliton formation with fast sweeps. (C) Basic operation procedure for adjustment of sweep parameters for generating solitons with fast sweeps. Red (blue) traces correspond to cases where the final state is effectively red (blue).

Our soliton generation technique (Fig. S2) addresses the unique challenge of fast thermal dynamics in compact SiN resonators (*5*, *11*), and our need to create solitons in specific ring resonators that have useful dispersive waves and low enough $f_0$ frequency for f-2f self-referencing. Fig S2 explores our technique in more detail. Our approach is similar to the open-loop slow laser tuning technique reported in Ref (*2*) but differs in the orders-of-magnitude faster rate of detuning control, which enables unprecedented fidelity of soliton generation (see Fig. 2C of main text). To illustrate the benefit of fast detuning control we first examine soliton dynamics for the case of slow detuning control for the SiN resonator used in this work (Fig. S2A). Here the laser is scanned over a range of >30 GHz and at a rate ≈ 500 GHz/s. The transition from the MI state to the soliton states with either N=3,2,1 pulses is accompanied by a precipitous drop in intracavity power resulting in significant cooling. As a result, the resonance frequency blue shifts towards its "cold-cavity" frequency at the thermal time scale of the resonator. Once the effective pump-resonator detuning exceeds the existence range (*i.e.: too red*), the soliton is lost within a few ns. The duration of the soliton steps in this regime of adiabatic detuning control is a complicated function of the temperature and intracavity power change and the extent of the pump-power-dependent soliton existence range (*11*), but are generally the same order of magnitude as the timescale of the thermal dynamics which are in the range of 100 ns – 5 µs for such resonators (*5*, *11*, *12*). The general absence of step durations that exceed 1 µs even when the scan speed is reduced by over an order of magnitude indicate that thermal relaxation is the dominant mechanism that determines the step duration in the adiabatic regime.

Fig S2B examines soliton dynamics in the opposite regime, where the pump laser is tuned faster than the thermal timescale. Here, the pump laser starts blue detuned and its frequency is decreased by 5 GHz in under 100 ns, reaching the effectively red detuned single soliton state at ~0 ns. Because the soliton state is accessed faster than the thermal response, the resonator is in equilibrium with the ambient environmental temperature. The sudden increase in intracavity power results in delayed 'thermal charging', which shifts the cavity resonance red and eventually causes the soliton to be lost because the effective detuning becomes too small (~14 ns in Fig. S2B). At this point, the comb traces back over the blue detuned MI states. A double exponential fit to the decaying tails on the blue detuned side, gives time constants of $\tau_1 \approx 100$ ns and $\tau_2 \approx 1.5$ µs. The faster of these is attributed with thermalization of the $\approx 1$ µm$^2$ mode volume and the slower with the thermalization of the resonator as a whole (*13*). A third timescale on the order of 100 µs – 1 ms is not evident in Fig. S2B but is attributed to thermalization of the entire chip (Fig. S1A).

Fig. S2C outlines how the general behavior illustrated in Fig. S2A and S2B are used to optimize the laser frequency ramp for successful generation of solitons. To stabilize the soliton state, the final value of the detuning is progressively increased to chase the thermally red-shifting resonance. Each of the five panels in Fig. S2C corresponds to a different final detuning which increases monotonically from panel I to panel V. In panels I, II, III, and IV, the increase in final detuning leads to an increased duration of the soliton state until a stable soliton is generated that can circulate indefinitely within the cavity (panel IV). Increasing the detuning past the existence range results in the soliton being lost within a few ns without retracing the blue detuned MI comb (panel V) similar to the traces in Fig. S2A.

A striking feature of Fig. 2C of the main text is that not only are solitons generated with high success but that the single soliton (N=1) is generated with overwhelming probability. In the standard LLE theory each soliton constituting a multisoliton state is identical and in the absence of thermal effects should have the same existence range. Recent investigations have established that the inclusion of thermal dynamics breaks this degeneracy, such that different numbers of solitons have distinct existence ranges (*6*). The resulting thermal shift of the resonance frequency can be approximated by $\Delta v_T \propto P_{CW} + NP_S$ where $P_{CW}$, $P_S$ are the power of the CW background and each of the N solitons respectively. An increase in soliton number N results in a corresponding increase of thermal charging and thus a different final endpoint of the detuning. We speculate that the combination of ultrafast ramps and the thermal charging phenomena from multiple solitons allows the single soliton to be uniquely selected for by control of the final laser detuning value. Radiofrequency control of our sweeps also likely helps with repeatability.

*Hysteretic behavior of dispersive wave emission and effects on comb parameters.*

Movie S1 provides additional information on the mode interaction mediated dispersive wave formation. Soliton spectra are recorded for optical frequencies above 176 THz as the pump laser frequency is decreased in steps of 500 MHz. Coinciding with the disappearance of the mode interaction near 205 THz, the dispersive wave at 300 THz undergoes a sudden increase in optical power by 20 dB.

Movie S1: Mode interaction mediated dispersive wave formation. Each frame corresponds to the pump laser frequency being decreased by 500 MHz. The sequence loops 5 times-twice at a rate of 14 fps, then once at a rate of 7 fps, then twice more at 14 fps.

This binary detuning dependent DW formation is also accompanied by other changes in SiN comb parameters. Figure S3 shows the variation of the optical power of the short wavelength dispersive wave at 300 THz (top), the comb line spacing, (middle) and offset frequency (bottom) as the pump frequency is varied. Evidently, the soliton can exist on either of two branches, corresponding to the strength of the 300 THz dispersive wave. The long wavelength dispersive wave at 137 THz is relatively unaffected during these changes in detuning. As the pump frequency is decreased past a certain threshold the soliton transitions from the lower branch to the upper branch (blue arrow in top panel of Fig. S2). In general, the lower branch (blue circles) can be accessed during the initial fast sweep to detuning $\delta_1$ and the upper branch (red circles) is accessed by a secondary slow sweep to detuning $\delta_2$ (see also Fig. 2D and 2E of main text). If the laser frequency is then increased the soliton will eventually transition from the upper branch down to the lower branch (red arrow in Fig. S2). Interestingly, the upward and downward transitions do not occur at the same frequency, and the soliton exhibits pronounced hysteresis with a bistable region that exists for nearly 1 GHz.

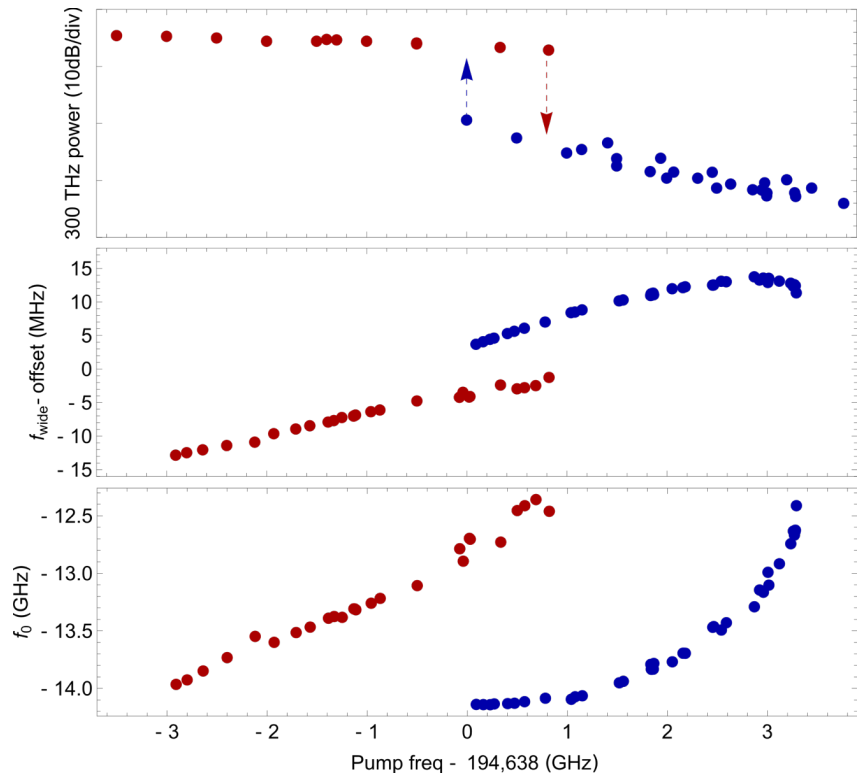

Figure S3: Hysteretic behavior of the short wavelength dispersive wave at 300 THz (top) and the simultaneous effect on the comb line spacing (middle) and offset frequency (bottom) as a function of the pump laser frequency. The blue and red points correspond to the states accessed at detuning $\delta_1$ and $\delta_2$.

## Supplemental Data

*Verifying coherence across comb.*

Figure S4 presents additional measurements of the μ= (-1,1,2,3) comb modes of the SiN comb. Data from Fig. 4A for the μ=0 mode is reproduced for comparison. The inaccuracy of each comb line is within the experimental uncertainty for the respective measurement.

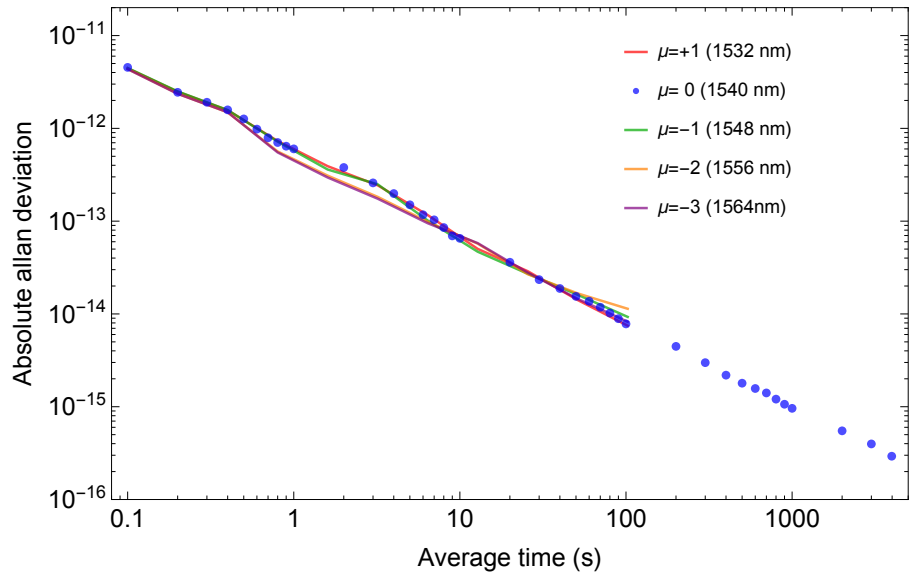

Figure S4 : Verifying coherence across comb for modes the µ= (-1,1,2,3). Data for µ=0 comb mode is reproduced from Fig. 4A for comparison.